\documentclass[preprint,aps,12pt,showpacs,nofootinbib,tightenlines]{revtex4}
\usepackage{amsmath}
\usepackage{amssymb}
\usepackage{epsfig}
\usepackage{graphicx}
\begin{document}

\newcommand{\beq}{\begin{eqnarray}}
\newcommand{\eeq}{\end{eqnarray}}
\newcommand{\tab}[1]{Table \ref{#1}}
\newcommand{\fig}[1]{Fig.\ref{#1}}
\newcommand{\real}{{\rm Re}\,}
\newcommand{\im}{{\rm Im}\,}
\newcommand{\non}{\nonumber\\ }

\title{Exclusive $B \to VV $ Decays and CP Violation in the General two-Higgs-doublet Model}
\author{Shou-Shan Bao}
\author{Fang Su}
\author{Yue-Liang Wu} \email{ylwu@itp.ac.cn}
\author{Ci Zhuang} \affiliation{ Kavli Institute for
Theoretical Physics China, Institute of theoretical physics\\
Chinese Academy of Science (KITPC/ITP-CAS), Beijing,100080,
P.R.China}
\date{\today}
\begin{abstract}
Using the general factorization approach, we present a detailed
investigation for the branching ratios, CP asymmetries and
longitudinal polarization fractions in all charmless hadronic $B \to
VV$ decays~(except for the pure annihilation processes) within the
most general two-Higgs-doublet model with spontaneous CP violation.
It is seen that such a new physics model only has very small
contributions to the branching ratios and longitudinal polarization
fractions. However, as the model has rich CP-violating sources, it
can lead to significant effects on the CP asymmetries, especially on
those of penguin-dominated decay modes, which provides good signals
for probing new physics beyond the SM in the future B-physics
experiments.
\end{abstract}

\pacs{12.60.Fr; 13.25.Hw; 11.30.Hv.} \maketitle

\newpage

\section{introduction}
During the recent years, tremendous progress in B physics has been
made through the fruitful interplay between theory and experiment.
The precise measurements of the $B$-meson decays can provide an
insight into very high energy scales via the indirect loop effects
of new physics beyond the standard model~(SM), which makes the study
of exclusive non-leptonic $B$-meson decays of great interest.

In the SM, the phenomenon of CP violation can be accommodated in an
efficient way through a complex phase entering the quark-mixing
matrix, which governs the strength of the charged-current
interactions of the quarks. This Kobayashi-Maskawa~(KM)~\cite{km}
mechanism of CP violation is the subject of detailed investigation
in these few decades. However, its origin remains unknown as it is
put into the standard model through the complex Yukawa couplings.
Moreover, the baryon asymmetry of the universe requires new sources
of CP violation. Many possible extensions of the SM in the Higgs
sector have been proposed~\cite{HS}, and it was suggested that CP
symmetry may break down spontaneously~\cite{Lee}. A consistent and
simple model, which provides a spontaneous CP violation mechanism,
has been constructed completely in a general two-Higgs-doublet
model~(2HDM)~\cite{YLW1,YLW2} without imposing the \textit{ad hoc}
discrete symmetry, which is now commonly called as type III 2HDM.
The type III 2HDM, which allows flavor-changing neutral
currents~(FCNCs) at tree level but suppressed by approximate $U(1)$
flavor symmetry, has attracted much more interests. It is known that
FCNCs are suppressed in low-energy experiments, especially for the
lighter two generation quarks. Thus, the type III 2HDM can be
parameterized in a way to satisfy the current experimental
constraints. On the other hand, constrains on the general 2HDM from
the neutral mesons mixing~($K^0-\bar{K}^0$, $D^0-\bar{D}^0$, and
$B^0-\bar{B}^0$)~\cite{Wolfenstein:1994jw, wumixings} and from the
radiative decays of bottom quark~\cite{b2gamma} have also been
studied in detail.

In recent years, there are many works about the $B$-meson decays
within the two-Higgs-doublet model. In Refs.~\cite{Xiao:2002mr,
B2PV}, the authors have studied the $B\to PP, PV$ decays~(with $P$
and $V$ denoting the pseudoscalar and vector mesons, respectively)
within the type III 2HDM. Since through the measurements of
magnitudes and phases of various helicity amplitudes, the charmless
hadronic $B\to VV$ decay modes can reveal more dynamics of exclusive
$B$ decays than $B\to PP$ and $B\to PV$ decays, in the present work
we are going to make a detailed study for $B\to VV$ decays within
the type III 2HDM by emphasizing on the new physics contributions.
It will be seen that this specific new physics has remarkable
effects on CP asymmetries, especially on the parameter $S_f$ for the
penguin-dominated decay modes. On the other hand, the new physics is
found to have very small contributions to the branching ratios and
the transverse polarizations. Furthermore, the polarization anomaly
observed in $B\to \rho K^*$ and $B\to \phi K^*$ modes can not be
improved in our current considered parameter spaces.

The paper is organized as follows: In section II, we first describe
the theoretical framework, including a brief introduction for the
two-Higgs-doublet model with spontaneous CP violation, the effective
Hamiltonian, as well as the decay amplitudes and CP violation
formulas, which are the basic tools to estimate the branching ratios
and CP asymmetries of $B$-meson decays. In section III, we list the
Wilson coefficients and the other relevant input parameters. Our
numerical predictions for the branching ratios, CP asymmetries and
longitudinal polarization fractions are presented in Section IV. Our
conclusions are presented in the last section.

\section{Theoretical Framework}

\subsection{Outline of the Two-Higgs-doublet Model}

Motivated solely from the origin of CP violation, a general
two-Higgs-doublet model with spontaneous CP violation~(type III
2HDM) has been shown to provide one of the simplest and attractive
models in understanding the origin and mechanism of CP violation at
the weak scale. In such a model, there exists more physical neutral
and charged Higgs bosons and rich CP violating sources from a single
CP phase of the vacuum. These new sources of CP violation can lead
to some significant phenomenological effects, which are promising to
be tested by the future $B$ factory and the LHCb experiments. In
this paper, we shall focus on the phenomenological applications of
the type III 2HDM on the two-body charmless hadronic $B\to VV$
decays.

The two complex Higgs doublets in the general 2HDM are generally
expressed as~\cite{YLW1,YLW2,LO,Kiers:1998ry}
\begin{equation}
\Phi_1=\left(\begin{array}{c} \phi_1^+\\ \phi_1^0
\end{array}\right), \, \, \, \Phi_2=\left(\begin{array}{c} \phi_2^+\\
\phi_2^0\end{array}\right).
\end{equation}
The corresponding Yukawa Lagrangian is given as
\begin{eqnarray}
  \mathcal{L}_{Y}
  &=&\eta_{ija}\bar{\psi}_{i,L}\tilde{\Phi}_aU_{j,R}+\xi_{ija}
  \bar{\psi}_{i,L}\Phi_aD_{j,R}+h.c.,
\end{eqnarray}
where the parameters $\eta_{ija}$ and $\xi_{ija}$ are real, so that
the lagrangian is CP invariant. After the symmetry is spontaneously
broken down
\begin{equation}
\langle\phi_1^0\rangle=v_1 e^{i \alpha_1}, \, \, \,
\langle\phi_2^0\rangle=v_2 e^{\alpha_2},
\end{equation}
and the Goldstone particles have been eaten, the physical Higgs
bosons are
\begin{equation}
H_1=\frac{1}{\sqrt{2}}\left( \begin{array}{c} 0\\ v+\phi_1^0
\end{array}\right), \, \, \, H_2=\frac{1}{\sqrt{2}}\left(
\begin{array}{c} H ^+\\ \phi_2^0+i \phi_3^0 \end{array}\right).
\end{equation}
where $H^{\pm}$ are the charged scalar mass eigenstates, ($\phi_1^0,
\phi_2^0, \phi_3^0$) are generally not the mass eigenstates but can
be expressed as linear combinations of the mass eigenstates ($H, h,
A$).

Then the Yukawa part of the Lagrangian for physical particles can be
written as
\begin{equation}
  \mathcal{L}_{Y}
  =\eta_{ij}^U\bar{\psi}_{i,L}\tilde{H}_1U_{j,R}+\eta_{ij}^D
  \bar{\psi}_{i,L}H_1D_{j,R}+\xi_{ij}^U\bar{\psi}_{i,L}\tilde{H}_2
  U_{j,R}+\xi_{ij}^D\bar{\psi}_{iL}H_2D_{j,R}+h.c.,
\end{equation}
where
\begin{eqnarray}
\eta_{ij}^U&= & \eta_{ij1}\cos\beta+\eta_{ij2}e^{-\delta}\sin\beta,
\nonumber\\
\xi_{ij}^U&= &-\eta_{ij1}e^{-\delta}\sin\beta+\eta_{ij2}\cos\beta,
\nonumber \\
\eta_{ij}^D&= & \xi_{ij1}\cos\beta+\xi_{ij2}e^{-\delta}\sin\beta,
\nonumber\\
\xi_{ij}^D&= &-\xi_{ij1}e^{-\delta}\sin\beta+\xi_{ij2}\cos\beta,
\end{eqnarray}
and these couplings $\eta^U, \eta^D, \xi^U, \xi^D$ are generally
complex, which means CP violation. According to the CKM mechanism,
after diagonalizing the fermion terms' couplings $\eta^U$ and
$\eta^D$, the other couplings become
\begin{eqnarray}
\label{Yukawa}
  \mathcal{L}_{Y}&=&\bar{U}_{i}\frac{m^U}{v}U_{R}(v+
  \phi_1^0)+\bar{D}_{L}\frac{m^D}{v}D_{R}(v+\phi_1^0)\nonumber\\
  &&+\bar{U}_{L}\tilde{\xi}^U U_{R}(\phi_2^0+i \phi_3^0)+\bar{D}_{L}
  \hat{\xi}^U U_{R}H^-\nonumber\\
  &&+\bar{U}_{L}\hat{\xi}^D D_{R}H^++\bar{D}_{L}\tilde{\xi}^DD_{R}
  (\phi_2^0+i \phi_3^0)+h.c,
\end{eqnarray}
with
\begin{eqnarray}
\tilde{\xi}^{U,D}&=&(V_L^{U,D})^{-1} \xi^{U,D}V_R^{U,D},\nonumber\\
\hat{\xi}^{U}&=&\tilde{\xi}^{U} V_{\mbox{CKM}},\nonumber\\
\hat{\xi}^D&=&V_{\mbox{CKM}}\tilde{\xi}^{D}.
\end{eqnarray}

The Yukawa couplings may be parameterized as following
\begin{equation}\label{yukawa}
\tilde{\xi}_{ij}=\lambda_{ij}\frac{\sqrt{m_i m_j}}{v}.
\end{equation}
with $v$ the vacuum expectation value $v=246$ GeV.

\subsection{Effective Hamiltonian and decay amplitudes of $B\to VV$ decays}

Using the operator product expansion and the renormalization group
equation, the low energy effective Hamiltonian for charmless
hadronic $B$-meson decays with $\Delta B=1$ can be written as
\begin{eqnarray}\label{eff}
 {\cal{H}}_{eff}= \frac{G_F}{\sqrt2}
 \sum_{p=u,c} \!
   V_{pb} V^*_{pq} \Big( C_1\,Q_1^p + C_2\,Q_2^p
   + \!\sum_{i=3,\dots, 16}\!\big[ C_i\,Q_i+ C_i^\prime\,Q_i^\prime \big]
    \Big) + \mbox{h.c.} \, ,
\end{eqnarray}
where $C_i(\mu)~(i=1,\dots, 16)$ are the Wilson coefficients that
can be calculated by perturbative theory, and $Q_i$ are the quark
and gluon effective operators, with $Q_{1-10}$ and $Q_{11-16}$
coming from the SM and from the type III 2HDM, respectively. Their
explicit forms are defined as follows~(taking $b\to sq\bar q$
transition as an example)~\cite{huang}
\begin{eqnarray}\label{operators}
&&Q_1 = (\bar{s} u)_{V-A} (\bar{u} b)_{V-A},\nonumber\\
&&Q_2 = (\bar{s}_i u_j)_{V-A} (\bar{u}_j b_i)_{V-A},\nonumber\\
&&Q_{3(5)} = (\bar{s} b)_{V-A}\sum_{q} (\bar{q} q)_{V-(+)A},\nonumber\\
&&Q_{4(6)} = (\bar{s}_i b_j)_{V-A}\sum_{q}(\bar{q}_j q_i)_{V-(+)A},
\nonumber\\
&&Q_{7(9)} =\frac{3}{2}(\bar{s} b)_{V-A}\sum_{q} e_{q}
(\bar{q} q)_{V+(-)A},\nonumber\\
&&Q_{8(10)} =\frac{3}{2}(\bar{s}_i b_j)_{V-A}\sum_{q}e_{q}
(\bar{q}_j q_i)_{V+(-)A},\nonumber\\
&&Q_{11(13)} = (\bar{s}b)_{S+P}\sum_q\frac{m_q\lambda_{qq}^{*}
(\lambda_{qq})}{m_b} (\bar{q} q)_{S-(+)P},\nonumber\\
&&Q_{12(14)} = (\bar{s}_i b_j)_{S+P}
\sum_q\,\frac{m_q\lambda_{qq}^{*}
(\lambda_{qq})}{m_b}(\bar q_j \,q_i)_{S-(+)P},\nonumber\\
&&Q_{15} = \bar s \,\sigma^{\mu\nu}(1+\gamma_5) b
\sum_q\,\frac{m_q\lambda_{qq}}{m_b}
\bar q \sigma_{\mu\nu}(1+\gamma_5)\,q,\nonumber\\
&&Q_{16} = \bar s_i \,\sigma^{\mu\nu}(1+\gamma_5) \,b_j \sum_q\,
\frac{m_q\lambda_{qq}}{m_b} \bar q_j\, \sigma_{\mu\nu}(1+\gamma_5)
\,q_i,
\end{eqnarray}
where $(\bar q_1 q_2)_{V\pm A}=\bar q_1\gamma_\mu(1\pm\gamma_5)q_2$
and $(\bar q_1 q_2)_{S\pm P}=\bar q_1(1\pm\gamma_5)q_2$, with $q
u,d,s,c,b$, and $e_{q}$ is the electric charge number of $q$ quark.
The operators $Q^\prime_i$ in Eq. (\ref{operators}) are obtained
from $Q_i$ via exchanging $L \leftrightarrow R$, and we shall
neglect their effects in our calculations for they are suppressed by
a factor $m_s/m_b$ in model III 2HDM. The Wilson coefficients
$C_i~(i=1,\dots,10)$ have been calculated at leading
order~(LO)~\cite{buchalla, paschos} and at next-to-leading
order~(NLO)~\cite{sm:nlo} in the SM and also at LO in
2HDM~\cite{LO}, while $C_i~(i=11,\dots,16)$ at LO can be found in
Refs.~\cite{huang, Dai}.

Having defined the effective Hamiltonian $H_{eff}$ in terms of the
four-quark operators $Q_i$, we can then proceed to calculate the
hadronic matrix elements with the generalized factorization
assumption~\cite{hyc1,hyc2,ali,lucd} based on the naive
factorization approach.

For two-body charmless hadronic $B\to VV$ decays, the decay
amplitude of the local four fermion operators is defined as
\begin{equation}
A_h\equiv \frac{G_F}{\sqrt{2}}\langle V_1(h_1)V_2(h_2)|(\bar
q_2q_3)_{V\pm A}(\bar b q_1)_{V-A} |B\rangle,
\end{equation}
where $h_1$ and $h_2$ are the helicities of the final-state vector
mesons $V_1$ and $V_2$ with four-momentum $p_1$ and $p_2$,
respectively. Since the $B$ meson has spin zero, in the rest frame
of $B$-meson system, the two vector mesons have the same helicity
due to helicity conservation. Therefore three polarization states
are possible in $B\to VV$ decays with one longitudinal~($L$) and two
transverse, corresponding to helicities $h=0$ and $h=\pm$~(here
$h_1=h_2=h$), respectively. We define the three helicity amplitudes
as follows
\begin{eqnarray}
A_0=A(B\to V_1(p_1,\epsilon_1^0) V_2(p_2, \epsilon_2^0)),\nonumber\\
A_{\pm}=A(B\to V_1(p_1,\epsilon_1^{\pm}) V_2(p_2,
\epsilon_2^{\pm})).
\end{eqnarray}

We choose the momentum $\vec{p}_2$ to be directed in the positive
$z$-direction in the $B$-meson rest frame, and the polarization
four-vectors of the light vector mesons such that in a frame where
both light mesons have large momentum along the $z$-axis. They are
given by
\begin{eqnarray}
\epsilon_1^{\pm \mu}&=&\epsilon_2^{\mp\mu}=(0,\pm1,i,0)/\sqrt{2} ,
\nonumber\\
\epsilon_{1,2}^{0\mu}&=&p_{1,2}^\mu/m_{1,2},
\end{eqnarray}
where $m_1$ and $m_2$ are the masses of $V_1$ and $V_2$ mesons,
respectively. Using the definitions for decay constants and form
factors~\cite{Beneke:2000wa}, the tree-level hadronic matrix
elements of the effective operators $Q_i$ can be decomposed as the
following two amplitudes
\begin{equation}\label{amplitude}
A_h=\mathcal{V}_h+\mathcal{T}_h,
\end{equation}
with
\begin{eqnarray}
\mathcal{V}_h &\equiv& \langle V_1(p_1,\epsilon_1^h)|V-A|B\rangle
\langle V_2(p_2,\epsilon_2^h)|V-A|0\rangle,\nonumber\\
\mathcal{T}_h &\equiv& \langle
V_1(p_1,\epsilon_1^h)|\sigma^{\mu\nu}(1+\gamma^5)|B\rangle \langle
V_2(p_2,\epsilon_2^h)|\sigma_{\mu\nu} (1+\gamma^5)|0\rangle.
\end{eqnarray}
Here, for simplicity, we have omitted the quark spinors in the
corresponding current operators in the above definitions. The three
polarization amplitudes for $\mathcal {V}^h$ and $\mathcal {T}^h$
can be further written as
\begin{eqnarray}
\label{factorization_fumula1}
&&\mathcal{V}_0=i f_{V_2}(m_B^2-m_1^2-m_2^2) A^{V_1}_0,\nonumber\\
\label{factorization_fumula2} &&\mathcal{V}_{\pm}=i f_{V_2} m_2
\left[A^{V_1}_1(m_1+m_B) \mp V^{V_1} \frac{2m_B |p_c|}
{m_B+m_1}\right],\nonumber\\ \label{factorization_fumula3}
&&\mathcal{T}_0=0,\nonumber\\ \label{factorization_fumula4}
&&\mathcal{T}_{\pm}=2 i f_{V_2}^{\perp}\bigg[ 2 T^{V_1}_1 m_B |p_c|
\mp T^{V_1}_2 (m_B^2-m_1^2)\bigg].
\end{eqnarray}

From the amplitude given by Eq.~(\ref{amplitude}), the branching
ratio for $B\to VV$ decays then reads
\begin{equation}
Br(B\to VV)=\frac{\tau_B|p_c|}{8\pi
m^2_B}\left(|A_0|^2+|A_+|^2+|A_-|^2\right),
\end{equation}
where $\tau_B$ is the lifetime of the $B$ meson, and $p_c$ is the
center of mass momentum of either final-state meson with
\begin{equation}
|p_c|=\frac{\sqrt{\left[m_B^2-(m_1+m_2)^2\right]
\left[m_B^2-(m_1-m_2)^2\right]}}{2 m_B}.
\end{equation}

In order to compare the relative size of the three different
helicity amplitudes, we can define the longitudinal polarization
fraction as
\begin{equation}
f_L=\frac{|A_0|^2}{|A_0|^2+|A_+|^2+|A_-|^2},
\end{equation}
which measures the relative strength of the longitudinally
polarization amplitude in a given decay mode.

\subsection{CP-violating asymmetries in $B\to VV$ decays}

Since there are abundant CP violation sources in the
two-Higgs-doublet model, it is also necessary and interesting for us
to discuss CP asymmetries in $B\to VV$ decays.

Firstly, for charged $B^{\pm}$-meson decays, there is only one
simple type of CP violating asymmetry, which detects direct CP
violation
\begin{eqnarray}
\mathcal{A}_{\mathcal{CP}}\equiv\frac{\Gamma(B^+\to
f^+)-\Gamma(B^-\to f^-)}{\Gamma(B^+\to f^+)+\Gamma(B^-\to f^-)}
\end{eqnarray}

For neutral $B$-meson decays, there is another type of CP violation
coming from the mixing between $B_q^0-\overline B_q^0$~(here $q=d$
or $s$)
\begin{eqnarray}
|B_q^0(t)\rangle&=&g_+(t)|B_q^0\rangle+\frac{q}{p}g_-(t)|\overline
B^0_q\rangle, \nonumber\\
|\overline B^0_q(t)\rangle&=&\frac{p}{q}g_-(t)|B_q^0\rangle+g_+
|\overline B^0_q\rangle.
\end{eqnarray}
In this case, there are in general four amplitudes which can be
expressed as~\cite{PheCP,WuCP,Yao:2006px}
\begin{eqnarray}\label{generala}
A_f=\langle f|H_{eff}|B_q^0\rangle &, & \overline A_f=\langle
f|H_{eff}|\overline B_q^0\rangle, \nonumber\\
\overline A_{\bar f}=\langle \bar f|H_{eff}|\overline
B_q^0\rangle&,& A_{\bar f} =\langle\bar f |H_{eff}|B_q^0\rangle.
\end{eqnarray}
For the $B_d-\overline B_d$ and $B_s-\overline B_s$ systems, the
following approximations can be made
\begin{eqnarray}\label{approx}
\mbox{both $B_d$ and $B_s$ systems}: \Big|\frac{q}{p}\Big|\sim1;
\qquad \mbox{only $B_d$ system}:  \Delta\Gamma\sim 0.
\end{eqnarray}
Using the decay amplitudes and the approximations listed in
Eqs.~(\ref{generala}) and (\ref{approx}), the time-dependent decay
probabilities for $B_d$ system can then be written as
\begin{eqnarray}
\Gamma(B_d^0(t)\to f)=\frac{|A_f|^2(1+|\lambda_f|^2)}{2}e^{-\Gamma
t}\left\{1+C_f \cos(\Delta mt)-S_f\sin(\Delta m t)\right\},
\nonumber\\
\Gamma(\overline B_d^0(t)\to
f)=\frac{|A_f|^2(1+|\lambda_f|^2)}{2}e^{-\Gamma t}\left\{1-C_f
\cos(\Delta mt)+S_f\sin(\Delta m t)\right\},\label{mixcpbd}
\end{eqnarray}
while for $B_s$ system, we have
\begin{eqnarray}
\Gamma(B_s^0(t)\to f)&=&\frac{|A_f|^2(1+|\lambda_f|^2)}{2}e^{-\Gamma
t} \Big[\cosh\Big(\frac{\Delta\Gamma t}{2}\Big)+D_f \sinh\Big(\frac
{\Delta\Gamma t}{2}\Big)\nonumber\\
&& \hspace{3.5cm} +C_f \cos(\Delta m
t)-S_f\sin(\Delta m t)\Big],\nonumber\\
\Gamma({\overline B_s^0}(t)\to f)&=&\frac{|{A}_f|^2(1+|{\lambda_
f|^2)}}{2}e^{-\Gamma t}\Big[\cosh\Big(\frac{\Delta\Gamma
t}{2}\Big)+D_f \sinh\Big(\frac{\Delta\Gamma t}{2}\Big)\nonumber\\
&& \hspace{3.5cm} -C_f \cos(\Delta{m} t)+S_f\sin(\Delta{m} t)\Big],\label{mixcpbs}
\end{eqnarray}
where $\Gamma$ is the average decay width, $\Delta\Gamma$ and
$\Delta m$ are the width and mass difference, respectively. The
other quantities are defined as
\begin{eqnarray}
\lambda_f\equiv\frac{q}{p}\frac{\bar{A}_f}{A_f},\hspace{1cm}
D_f\equiv\frac{2\mbox{Re}(\lambda_f)}{1+|\lambda_f|^2},\nonumber\\
C_f\equiv\frac{1-|\lambda_f|^2}{1+|\lambda_f|^2},\hspace{1cm}
S_f\equiv\frac{2\mbox{Im}(\lambda_f)}{1+|\lambda_f|^2}.
\end{eqnarray}

From Eqs.(\ref{mixcpbd}) and (\ref{mixcpbs}), we can get:
\begin{eqnarray}
  \mathcal{A_{CP}}(B_d\to f)&=&-C_f \cos \Delta m t+S_f \sin\Delta m t,\nonumber\\
  \mathcal{A_{CP}}(B_s\to f)&=&\frac{-C_f\cos \Delta mt+S_f\sin \Delta mt}{\cosh\left(\frac{\Delta \Gamma t}{2}\right)+D_f\sinh\left(\frac{\Delta \Gamma t}{2}\right)}.
\end{eqnarray}

\section{Input parameters}

The theoretical predictions in our calculations depend on many input
parameters, such as the Wilson coefficients, the CKM matrix
elements, the hadronic parameters, and so on. Here we present all
the relevant input parameters as follows.

It has been shown from $B_{d,s}^0-\overline B_{d,s}^0$ mixings that
the parameters $|\lambda_{cc}|$ and $|\lambda_{ss}|$ in
Eq.~(\ref{operators}) can reach to be around 100~\cite{mphya20},
while their phases are not well constrained. In our present work we
simply fix the phases to be $\pi/4$, and this choice will not cause
any trouble in our numerical results. For the parameters
$\lambda_{tt}$ and $\lambda_{bb}$, the constraints come mainly from
the experiments for $B-\bar{B}$ mixing, $\Gamma(b\to s \gamma)$,
$\Gamma(b\to c\tau\bar{\nu}_\tau$), $\rho_0$, $R_b$, $B\to P V$, and
the electric dipole moments~(EDMS) of the electron and
neutron~\cite{Atwood,huang,LO,Dai,B2PV}. Based on the above
analyses, we choose the following three typical parameter spaces
which are allowed by the present experiments and have been adopted
for the $B\to PV $ decays\cite{B2PV}
\begin{eqnarray*}
\mbox{Case\quad A}:\quad |\lambda_{tt}|&=&0.15; \quad |\lambda_{bb}|=50,\\
\mbox{Case\quad B}:\quad |\lambda_{tt}|&=&0.3;\ \quad |\lambda_{bb}|=30,\\
\mbox{Case\quad C}:\quad |\lambda_{tt}|&=&0.03; \quad
|\lambda_{bb}|=100,
\end{eqnarray*}
and $\theta_{tt}+\theta_{bb}=\pi/2$. For the Higgs masses and the
Wilson coefficients of $C_{1,\dots,10}$ corresponding to the SM, we
use the results listed in the paper~\cite{B2PV}, while for the
Wilson coefficients in the type III 2HDM, we redefine them as
$\tilde{C}_{11,\dots,16}= \frac{m_s\lambda^{(*)}_{ss}}{m_b}
C_{11,\dots,16}$ in order to compare the contributions from those
operators in SM, here the factor $\frac{m_s\lambda^{(*)}_{ss}}{m_b}$
is associated with the operators in 2HDM, the numerical values for
$\tilde{C}_{11,\dots,16}$ are listed in Table \ref{wilsonc}.

\begin{table}[ht]
\begin{center}
\caption{The Wilson coefficients $\tilde{C}_{11,\dots,16}=
\frac{m_s\lambda^{(*)}_{ss}}{m_b} C_{11,\dots,16}$ in $b \to s$
transition at $\mu =m_b=4.2~\rm{GeV}$ in 2HDM.} \label{wilsonc}
\doublerulesep 0.8pt \tabcolsep 0.15in\vspace{0.2cm}
\begin{tabular} {c c c c c }
\hline\hline
 & Case A & Case B & Case C \\
\hline

$\tilde{C}_{11} $&$-0.0085+0.012 i$&$-0.0085+0.018 i$&$-0.010+0.012 i$\\

$\tilde{C}_{12} $ &$0$&$0$&$0$\\

$\tilde{C}_{13} $ &$-0.0030-0.0049 i$&$-0.0052-0.0069 i$&$-0.0029-0.0052 i$\\

$\tilde{C}_{14} $ &$-0.000060-0.00010 i$&$-0.00011-0.00014 i$&$-0.000059-0.00010 i$\\

$\tilde{C}_{15} $ &$0.000033+0.000055 i$&$0.000058+0.000078 i$&$0.000032+0.000059 i$\\

$\tilde{C}_{16} $&$-0.00010-0.00017 i$&$-0.00018-0.00024 i$&$-0.0001-0.00018 i$\\
\hline\hline
\end{tabular}
\end{center}
\end{table}

As for the CKM matrix elements, we shall use the Wolfenstein
parametrization~\cite{ckm} with the values~\cite{Yao:2006px}:
$A=0.8533\pm0.0512$, $\lambda=0.2200\pm0.0026$, $\bar
\rho=0.20\pm0.09$, and $\bar \eta=0.33\pm0.05$.

For the hadronic parameters, the decay constants, and the form
factors, we list them in Tables.~\ref{input} and \ref{formfactors},
respectively.

\begin{table}[ht]
\begin{center}
\caption{The hadronic input parameters~\cite{Yao:2006px} and the
decay constants taken from the QCD sum rules~\cite{ball3} and
Lattice theory~\cite{lattice}.} \label{input} \doublerulesep 0.8pt
\tabcolsep 0.10in \vspace{0.2cm}
\begin{tabular}{cccccc} \hline\hline
$\tau_{B^\pm}$&$\tau_{B_d}$ & $\tau_{B_s}$&$M_{B_d}$ &$M_{B_s}$
&$m_b$ \\ \hline

$1.638$ps &$1.528$ps&$1.472$ps &$5.28$GeV &$5.37$GeV &$4.2$GeV\\

$m_t$&$m_u$&$m_d$&$m_c$&$m_s$&$m_{\rho^0}$\\

$174$GeV&$3.2$MeV&$6.4$MeV&$1.1$GeV&$0.105$GeV&$0.77$GeV\\

$m_{\rho^{\pm}}$&$m_{\omega}$&$m_{\phi}$&$m_{K^{*\pm}}$&$m_{K^{*0}}$
&$\Lambda_{QCD}$\\

$0.77$GeV&$0.782$GeV&$1.02$GeV&$0.892$GeV&$0.896$GeV&$225$MeV\\

$f_{\rho}$&$f_{\omega}$&$f_{K^*}$&$f_{\phi}$&$f_{\rho}^T$
&$f_{\omega}^T$\\

$0.205$GeV&$0.195$GeV&$0.217$GeV&$0.231$GeV&$0.147$GeV&$0.133$GeV\\

$f_{K^*}^T$&$f_{\phi}^T$& & \\

$0.156$GeV&$0.183$GeV\\ \hline\hline

\end{tabular}
\end{center}
\end{table}

\begin{table}[htbp]
\begin{center}
\caption{The relevant $B\to V$ transition form factors at $q^2=0$
taken from the light-cone sum
rules~(LCSR)~\cite{prd71014029,Wu:2006rd}.}\label{formfactors}
\doublerulesep 0.8pt \tabcolsep 0.15in \vspace{0.2cm}
\begin{tabular}{c c c c c c c }\hline\hline
decay channel& $V$& $A_0$ & $A_1$ & $A_2$ & $T_1$ &$T_3$
\\\hline
$B \to \rho$ & 0.323 &0.303 & 0.242 &0.221 &0.267 &0.176 \\
$B \to \omega$ &0.293 &0.281 & 0.219 &0.198 &0.242 &0.155 \\
$B\to K^*$ &0.411 &0.374 &0.292 &0.259 &0.333 &0.202 \\
$B_s \to \bar{K}^*$ & 0.311 & 0.363 &  0.233 &0.181 &0.26 & 0.136\\
$B_s \to \phi$&0.434 &  0.474 & 0.311 &0.234 & 0.349 &0.175 \\
\hline\hline
\end{tabular}
\end{center}
\end{table}

\section{Numerical results and discussions}

In this section, we shall classify the $28$ channels of $B^+$, $B^0$
and $B_s$ decays into two light vector mesons according to the
reliability of the calculation for various observables, which is
motivated by the dominated contributing operators. We shall give our
predictions for the branching ratios, the CP asymmetries, and the
longitudinal polarization fractions both in the SM and in the 2HDM.
Comparisons with the current experiment data, if possible, are also
made.

Before moving to the detailed discussions, some general observations
of new physics effects on $B\to VV$ decays should be made. As can be
seen from Eqs.~(\ref{yukawa}) and (\ref{operators}), the
contributions of new physics operators $O_{11,\dots,16}$ are always
proportional to the factor ${m_q}/{v}$. Thus, they are severely
suppressed for the first generation quarks. In this case, for $B\to
\rho K^*, \omega K^*, \rho \rho, \omega \rho, \omega \omega$ and
$B_s\to \rho K^*, \omega K^*, K^* K^*, \rho \phi , \omega \phi$
decay channels, we can safely ignore the contributions from  those
new operators. Note that the new physics still has effects on the
Wilson Coefficients $C_{1-10}$. On the other hand, for $B\to \phi
K^*, \phi \rho, \phi \omega$ and $B_s \to \phi K^*, \phi \phi$ decay
channels, since these are all induced by $b\to q s\bar s$~($q=d, s$)
transitions, we could not ignore the new operators' contributions
any more in this case. In the general factorization approach, it is
impossible to produce a vector meson via the scalar and/or
pseudoscalar currents from the vacuum state, and hence the new
operators $Q_{11}$ and $Q_{13}$ have no contributions to $B\to VV$
decays. Moreover, from the results listed in Table \ref{wilsonc}, it
can be seen that all the contributing new operators $Q_{12, 14, 15,
16}$ have only very small~(even zero) Wilson coefficients. It is
therefore expected that the new physics will have very small effects
on the branching ratios and transverse amplitudes~(hence on the
transverse polarization fractions) of $B\to VV$ decays.

\subsection{CP-averaged branching ratios and direct CP violation}

According to different decay modes, we shall give our predictions
for the branching ratios and direct CP violations one by one.

(i), color-allowed tree-dominated decays. Our predictions for the
CP-averaged branching ratios and the direct CP asymmetries are
presented in Table~\ref{tree}. From the numerical results, we can
see that the branching ratios are all at $10^{-5}$ order, and the
direct CP asymmetries are all very small since the penguin amplitude
contributions are much smaller than the ones from the tree diagrams.
Most predictions within the SM are consistent with the current
experiment data, and the new physics has very small effects on these
types of decays.
\begin{table}[htbp]
\caption{The CP-averaged branching ratios~(in unit of $10^{-6}$)
~(first line) and the direct CP violations~(second line) for the
color-allowed tree-dominant processes both in the SM and in the type
III 2HDM. Case A-C stand for the three different parameter spaces
listed in Section III.}\label{tree}
\begin{center}
\doublerulesep 0.8pt \tabcolsep 0.15in
\begin{tabular}{lcccccc}\hline \hline
Decay modes & Case A &Case B &Case C &SM &Exp. \\
 \hline\hline
 $B^+ \to \rho^+ \rho^{0}$&14.59&14.59&14.59&15.53&18.2$\pm$3.0\\
  &-0.004&-0.004&-0.004&-0.002&-0.08$\pm$0.13\\
$B^0 \to \rho^+ \rho^-$&26.33&25.93&26.73&27.49&$24.2^{+3.1}_{-3.2}$\\
&-0.043&-0.043&-0.042&-0.035&\\
$B^+ \to \rho^+\omega$&12.66&12.47&12.85&13.97&$10.6^{+2.6}_{-2.3}$\\
 &-0.042&-0.043&-0.042&-0.034&0.04$\pm$0.18\\
$B_s \to \rho^+ K^{*-}$ &36.88  &36.32 &37.44  &38.50 &\\
&-0.043 & -0.043  & -0.042 &-0.035 &\\
  \hline\hline
\end{tabular}
\end{center}
\end{table}

(ii), color-suppressed tree-dominated decays. The numerical results
are given in Table~\ref{ctree}, it is interesting to note that the
branching ratios will generally become smaller after including the
new physics contributions except for the $B\to \rho^0\rho^0$ mode.
Furthermore, there are big direct CP violations in these decay
processes except for the $B^+\to \rho^0\omega$ mode, and the new
physics has more effects on the direct CP asymmetries than on the
branching ratios through the Wilson coefficient functions, although
there are no new operator contributions to the hadronic matrix
elements in this type decays within our approximations. Compared to
Case A and Case C, Case B has the biggest corrections to the CP
asymmetries of the SM.
\begin{table}[htbp]
\caption{The same as Table~\ref{tree} but for color-suppressed
tree-dominant processes.}\label{ctree}
\begin{center}
\doublerulesep 0.8pt \tabcolsep 0.15in
\begin{tabular}{lcccccc}\hline \hline
Decay modes & Case A &Case B &Case C &SM &Exp. \\
\hline\hline
$B^0 \to \rho^0 \rho^0$&0.0814&0.0897&0.0754&0.065&0.86$\pm$0.28\\
&0.176&0.218&0.119&0.153&\\
$B^+ \to \omega \omega$&0.112&0.110&0.115&0.160&$<$4.0\\
&-0.117&-0.088&-0.144&-0.207&\\
$B_s \to \rho^0 \bar{K}^{*0}$&0.081  &0.090 &0.073  &0.092 &$<7.67\times 10^{-4}$\\
&0.176 & 0.218 & 0.119 & 0.153  &\\
$B_s \to \omega \bar{K}^{*0}$&0.183 &0.180 &0.187 &0.262 &\\
 &-0.167 & -0.088 & -0.144 & -0.207 &\\
$B^+ \to \rho^0 \omega$&0.024&0.024&0.024&0.076&$<$1.5\\
 &-0.063&-0.063&-0.063&-0.035&\\
  \hline\hline
\end{tabular}
\end{center}
\end{table}

(iii), penguin-dominated decays. We may divide such decays into two
types: $\Delta S=1$ and $\Delta D=1$ decay modes. They are
corresponding to the upper and the lower parts in
Table~\ref{penguin}, respectively. From the numerical results, we
can see that all the eleven $\Delta S=1$ decay modes have branching
ratios up to $10^{-6}$ or even to $10^{-5}$ order, since they
involve the relative large CKM matrix elements $V^*_{ts}$, while the
$\Delta D=1$ ones have much smaller branching ratios of order of
$10^{-7}$ due to the smaller CKM matrix elements $V^*_{td}$. For
$B\to \omega K^*$ and $B_s \to \phi\phi$ decay modes, our
predictions for the branching ratios with including the new operator
contributions have similar results as the ones within the SM, which,
however, are not quite consistent with the current experimental
data; the numerical results for $B\to \omega K^{\ast}$ modes are
larger than the current experiment limit, and the prediction for
$B_s\to \phi\phi$ is about two times larger than the present data.
For the other decay modes, our predictions for the branching ratios
are in general agreement with the data. As for the direct CP
asymmetries, there are big CP violations in some decay modes, and
the new physics can lead to remarkable effects. Our predictions are
consistent to the data in all these decay modes.
\begin{table}[htbp]
\caption{The same as Table~\ref{tree} but for the penguin-dominated
decay modes. The upper and the lower parts correspond to $\Delta
S=1$ and $\Delta D=1$ processes, respectively.}\label{penguin}
\begin{center}
\doublerulesep 0.8pt \tabcolsep 0.15in
\begin{tabular}{l cccccc}\hline \hline
Decay modes & Case A &Case B &Case C &SM &Exp. \\
\hline\hline
$B^+ \to \rho^+ K^{*0}$&7.169&7.409&7.027&7.287&9.2$\pm$1.5\\
 &0.084&0.117&0.049&0.018&-0.01$\pm$0.16\\
$B^+ \to \rho^0 K^{*+}$&5.853&6.229&5.526&5.575&$<$6.1\\
 &0.184&0.196&0.169&0.122&$0.20^{+0.32}_{-0.29}$\\
$B^0 \to \rho^0 K^{*0}$&6.396&6.513&6.324&6.245&5.6$\pm$1.6\\
 &0.054&0.073&0.033&0.018&0.09$\pm$0.19\\
$B^0 \to \rho^- K^{*+}$&6.046&6.738&5.445&5.571&$<$12\\
 &0.295&0.301&0.283&0.199&\\
$B^0 \to \omega K^{*0}$&3.412&3.513&3.351&3.498&$<$2.7\\
 &0.078&0.107&0.048&0.024&\\
$B^+ \to \omega K^{*+}$&3.247&3.5697&2.965&3.123&$<$3.4\\
 &0.265&0.274&0.251&0.176&\\
$B^0 \to \phi K^{*0}$&9.276&9.704&9.221&9.318&9.5$\pm$0.8\\
 &0.045&0.081&-0.002&0.020&-0.01$\pm0.06$\\
$B^+ \to \phi K^{*+}$&9.867&10.32&9.775&9.979&10.0$\pm$1.1\\
 &0.039&0.074&-0.013&0.020&-0.01$\pm$0.08\\
 $B_s \to \phi \phi$&28.99 &30.34 &28.64 &28.85 &$14^{+8}_{-7}\times 10^{-6}$\\
 &0.054 & 0.089 &0.006 & 0.020 &\\
$B_s \to \bar{K}^{*0} K^{*0}$&9.303 &9.614 &9.118 &9.456 &$<1.681\times 10^{-3}$\\
 &0.084 & 0.117 & 0.049 & 0.018 &\\
$B_s \to K^{*+} K^{*-}$ &8.404 &9.366  &7.569 &7.744 &\\
 &0.295 & 0.302 & 0.283 & 0.199  &\\
\hline
$B^0 \to \bar{K}^{*0} K^{*0}$&0.410&0.420&0.413&0.408&$0.49^{+0.17}_{-0.14}$\\
 &-0.092&-0.061&-0.133&-0.145&\\
$B^+ \to K^{*+} K^{*0}$&0.439&0.450&0.443&0.437&$<2.2$\\
 &-0.092&-0.061&-0.133&-0.145&\\
 $B_s \to \phi \bar{K}^{*0}$&0.517 &0.532 &0.521 &0.526 &$<1.013\times 10^{-3}$\\
 & -0.094 & -0.056 & -0.145 & -0.161 &\\
\hline\hline
\end{tabular}
\end{center}
\end{table}

(iv), electroweak penguin or QCD flavor singlet dominated decays. As
can be seen from Table~\ref{ewpenguin}, this type of decays are
expected to have smaller branching ratios due to the large
cancelations among the different Wilson coefficients. Although there
are new operator contributions in $B\to \rho \phi$ and $\omega \phi$
decay modes, the predicted branching ratios are still small. The
direct CP asymmetries for these decays are all small, and the new
physics effects on these observables are not prominent. Due to the
lack of accurate experimental data, we couldn't compare our
predictions with the data yet.
\begin{table}[htbp]
\caption{The same as Talbe~\ref{tree} but for the electroweak
penguin or QCD flavor singlet dominated decays.}\label{ewpenguin}
\begin{center}
\doublerulesep 0.8pt \tabcolsep 0.15in
\begin{tabular}{lcccccc}\hline \hline
Decay modes & Case A &Case B &Case C &SM &Exp. \\
\hline\hline
$B^+ \to \rho^+ \phi$&0.0054&0.0054&0.0054&0.0043&$<16$\\
 &-0.011&-0.011&-0.011&-0.014&\\
$B^0 \to \rho^0 \phi$&0.0025&0.0025&0.0025&0.0020&$<$13\\
 &-0.011&-0.011&-0.011&-0.014&\\
$B^0 \to \omega \phi$&0.0022&0.0022&0.0022&0.0017&$<1.2$\\
 &-0.011&-0.011&-0.011&-0.014&\\

$B_s \to \rho^0 \phi$&0.796 &0.796 &0.796 &0.687 &$<6.17\times 10^{-4}$\\
 &0.0048 & 0.0048 &0.0048 & 0.0039 &\\
$B_s \to \phi \omega$   &0.038  &0.038 &0.038 &0.045 &\\
 &0.020 & 0.020 &0.020 & 0.018  &\\
\hline\hline
\end{tabular}
\end{center}
\end{table}

(v), the pure annihilation decays. Only six decays belong to this
class, namely $B^0\to K^{*+}K^{*-}$, $B^0\to\phi\phi$,
$B_s\to\rho^+\rho^-$, $B_s\to \rho^0\rho^0$, $B_s\to \rho^0\omega$,
and $B_s\to \omega\omega$. Due to the lack of the information for
the $V_1\to V_2$ transition form factor at large momentum transfers,
we shall not consider them in details in this paper.

\subsection{Time-dependent CP violating parameters $C_f$, $S_f$ and $D_f$}

Since there are abundant CP violating sources in type III 2HDM, it
is expected that there are relatively large CP violations in 2HDM
than in the SM. Using the relevant formulas given in section II, we
can predict the time-dependent CP asymmetries in neutral $B_d$ and
$B_s$ decays, with the numerical results given in Tables~\ref{Bd}
and \ref{Bs}, respectively.

From these two tables, it is seen that, for $B^0\to \rho^+\rho^-,
\rho^0 \phi$ and $\omega \phi$ decay modes, the new physics has
hardly any effects on the parameters $C_f$ and $S_f$, even though
there are new operators contributions in $B^0\to \rho^0 \phi$ and
$\omega \phi$ decay modes. On the other hand, the new physics has
remarkable effects on the other decay modes, especially on $B^0\to
\omega \omega$ one~(for this mode the new physics can even change
the sign of the parameter $S_f$). Furthermore, different parameter
spaces also have remarkable effects on these CP violation
parameters.

For $B_s$ system, there are new operator contributions only in
$B_s\to \phi\phi$ mode. As is expected, the new physics has
remarkable influence on the parameters $C_f$, $S_f$, and $D_f$. For
the other four decay modes, although there are no new operator
contributions, the new physics still has big effects on the
parameter $S_f$, but small effects on $C_f$ and $D_f$.
\begin{table}[htbp]
\caption{The time-dependent CP asymmetry parameters $C_f$~(first
line) and $S_f$~(second line) for $B_d$ decays both in the SM and in
the type III 2HDM. Case A-C stand for the three different parameter
spaces listed in Section III.}\label{Bd}
\begin{center}
\doublerulesep 0.8pt \tabcolsep 0.15in
\begin{tabular}{lccccc}\hline \hline
Decay modes & Case A &Case B &Case C &SM  \\
\hline\hline
$B^0 \to \rho^+ \rho^-$&0.043&0.043&0.042&0.035\\
 &-0.95&-0.95&-0.95&-0.95\\
$B^0 \to \rho^0 \rho^0$&-0.18&-0.22&-0.12&-0.15\\
 &0.97&0.92&0.99&0.89\\
$B^0 \to \omega \rho^0$&0.063&0.063&0.063&0.029\\
 &-0.61&-0.61&-0.62&-0.97\\
$B^0 \to \phi \rho^0$&0.011&0.011&0.011&0.014\\
 &0.70&0.70&0.70&0.70\\
$B^0 \to \omega \phi$&0.011&0.011&0.011&0.014\\
 &0.70&0.70&0.70&0.70\\
$B^0 \to \omega \omega$&0.12&0.09&0.14&0.21\\
 &0.53&0.65&0.40&-0.18\\
$B^0 \to K^{*0} \bar{K}^{*0}$&0.092&0.061&0.13&0.15\\
 &0.85&0.92&0.75&0.57\\
\hline\hline
\end{tabular}
\end{center}
\end{table}

\begin{table}[htbp]
\caption{The time-dependent CP asymmetry parameters $C_f$~(first
line), $S_f$~(second line), and $D_f$~(third line) for $B_s$ decays
both in the SM and in the type III 2HDM }\label{Bs}
\begin{center}
\doublerulesep 0.8pt \tabcolsep 0.15in
\begin{tabular}{lccccc}\hline \hline
Decay modes & Case A &Case B &Case C &SM  \\
\hline\hline
$B_s \to \phi \rho^0$&-0.005&-0.005&-0.005&-0.004\\
 &0.052&0.052&0.052&0.14\\
 &0.99&0.99&0.99&0.99\\
$B_s \to \phi \omega$&-0.020&-0.020&-0.020&-0.018\\
 &0.23&0.23&0.23&0.49\\
 &0.97&0.97&0.97&0.87\\
$B_s \to \phi \phi$&-0.054&-0.090&-0.060&-0.020\\
 &0.33&0.49&0.14&-0.004\\
 &0.94&0.87&0.99&1.0\\
$B_s \to K^{*+} K^{*-}$&-0.30&-0.30&-0.28&-0.20\\
 &0.92&0.95&0.88&0.79\\
 &0.25&0.12&0.39&0.57\\
$B_s \to \bar{K}^{*0} K^{*0}$&-0.085&-0.12&-0.049&-0.018\\
 &0.31&0.45&0.15&-0.003\\
 &0.95&0.88&0.99&1.0\\
\hline\hline
\end{tabular}
\end{center}
\end{table}

\subsection{The polarization in $B\to \rho K^*$ and $\phi K^*$ decays}

Motivated by the polarization anomaly observed by the
BarBar~\cite{babar}, Belle~\cite{belle} and CDF~\cite{cdf}
experiments, we shall study the polarization in $B\to VV$ decays,
especially in $B\to \rho K^*$ and $\phi K^*$ decays in this section.

One important point that should be noted is that the predictions for
the branching ratios of $B\to \rho K^*$ and $\phi K^*$ modes are
well consistent with the experiment data, which means that if we
want to solve the observed polarization anomaly, we need to find
some way to reduce the longitudinal amplitude and enhance transverse
ones simultaneously. Many studies have been made to try to provide
possible resolutions to the anomaly both within the
SM~\cite{Kagan:2004uw,Li:2004ti,Colangelo:2004rd,Beneke:2006hg} and
in various new physics
models~\cite{Yang:2004pm,Chen:2005mka,Chang:2006dh}. Here we only
concentrate on the longitudinal polarization fraction and the main
results are listed in Table~\ref{polarization}.

It is noted that the polarization anomaly could be well resolved by
introducing the tensor operators
$O_{T1}=\bar{s}\sigma^{\mu\nu}(1+\gamma^5) b \, \bar{s}
\sigma_{\mu\nu}(1+\gamma_5) s$ and
$O_{T8}=\bar{s}_i\sigma^{\mu\nu}(1+\gamma^5) b_j \, \bar{s}_j
\sigma_{\mu\nu}(1+\gamma_5) s_i$ in Ref.~\cite{Chang:2006dh}. It is
interesting to see that these two operators have similar forms as
$Q_{15}$ and $Q_{16}$ in Eq.~(\ref{operators}). However, from the
numerical results given by Table~\ref{polarization}, we can see that
the predicted longitudinal polarization fraction $f_L$ for these
decay modes in the type III 2HDM is almost the same as the one
within the SM. Although there are new operator contributions in
$B\to \phi K^*$ modes, we still can not resolve the polarization
anomaly observed in this decay mode. This is due to the fact that
the strength of new operators in 2HDM is severely suppressed by the
factor $m_q\lambda_{qq}/m_b$. Moreover, as has already been
mentioned in the beginning of this section, the Wilson coefficients
of these new operators are very small, which also result in the
small effects on the transverse amplitudes.

\begin{table}[htpb]
\caption{The longitudinal polarization fractions $f_L$ for $B\to
\rho K^*$ and $\phi K^*$ decay modes. Case A-C stand for the three
different parameter spaces in the type III
2HDM.}\label{polarization}
\begin{center}
\doublerulesep 0.8pt \tabcolsep 0.15in
\begin{tabular}{lccccc}\hline \hline
Decay modes &SM & Case A &Case B &Case C  &Exp. \\
\hline\hline
$B^+ \to \rho^+ K^{*0}$&0.91&0.91&0.91&0.91&$0.48\pm0.08$\\

$B^0 \to \rho K^{*0}$&0.95&0.95&0.93&0.93&$0.57\pm0.12$\\

$B^+ \to \phi K^*$&0.89&0.89&0.89&0.89&$0.50\pm0.05$\\

$B^0 \to \phi K^{*0}$&0.89&0.89&0.89&0.89&$0.491\pm0.032$\\

\hline\hline
\end{tabular}
\end{center}
\end{table}

For the other $B\to VV$ decay modes, the predictions for
longitudinal polarization fractions are always about $0.90\sim0.95$.
For simplify, we shall not list the results in details anymore.

In conclusion, adopting the current parameter spaces and with the
general factorization method, we could not resolve the polarization
anomaly observed in $B\to \rho K^*$ and $\phi K^*$ modes within the
SM and 2HDM.

\section{conclusions}

Using the general factorization approach, we have studied all the
$B\to VV$ decay modes except for pure annihilation decay channels
both within the SM and in the two-Higgs-doublet model. From the
numerical results given in the previous section, we can see that:
for the branching ratios, our predictions are generally well
consistent with the current experimental data expect for the $B_s\to
\phi\phi$ decay mode, and the new physics has margin or even
negligible effects on this observable. However, the new physics can
give remarkable contributions to the CP asymmetry parameters $C_f$
and $S_f$, especially to $S_f$ in the penguin-dominated decay modes.
Unfortunately, our predictions for the longitudinal polarization
fractions of $B\to \rho K^*$ and $\phi K^*$ decay modes in 2HDM are
still as large as the ones in the SM, which are much larger than the
experimental data. Some new mechanisms may be needed to improve
those discrepancies.

For simplicity, in this paper we have neglected the contributions
from annihilation and exchange diagrams, although they may play a
significant rule in some decay channels. In our numerical
calculations, we have only considered three possible parameter
spaces for the type III 2HDM. Also we have totally neglected the
first generation Yukawa couplings and the off-diagonal matrix
elements of the Yukawa coupling matrix, in order to eliminate the
FCNC at tree level. However, it is possible that the FCNC involving
the third generation quarks still exists at tree level, making the
constraints less stronger. In a word, we do not exclude the
possibility to improve the predictions by using the other
factorization methods with the annihilation and exchange diagram
contributions included, by choosing other parameters spaces, or even
by introducing additional fourth-generation quarks~\cite{Wu:2004kr}.

In conclusion, we have shown that the new Higgs bosons in the type
III 2HDM with spontaneous CP violation can have significant effects
on some charmless $B\to VV$ decays, especially for the
penguin-dominated decay modes, which can be used as good signals to
test the SM and to explore new physics from more precise
measurements in the future $B$-factory experiments.

\begin{acknowledgments}
This work was supported in part by the National Science Foundation
of China (NSFC) under the grant 10475105, 10491306, and the Project
of Knowledge Innovation Program (PKIP) of Chinese Academy of
Sciences.
\end{acknowledgments}


\end{document}